\documentclass[preprint,12pt]{elsarticle}

\usepackage{bm}

\usepackage{amssymb}

\newcommand{\bea}{\begin{eqnarray}}
\newcommand{\eea}{\end{eqnarray}}
\newcommand{\bel}[1]{\begin{eqnarray}\label{#1}}
\newcommand{\eel}{\end{eqnarray}}

\def\LB{\left(}
\def\RB{\right)}

\newcommand{\EQ}[1]{Eq.~(\ref{#1})}
\newcommand{\EQn}[1]{(\ref{#1})}

\newcommand{\EQSM}[2]{Eqs.~(\ref{#1})--(\ref{#2})}
\newcommand{\EQSMn}[2]{(\ref{#1})--(\ref{#2})}

\newcommand{\CIT}[1]{Ref.~\citep{#1}} 
\newcommand{\CITn}[1]{\citep{#1}} 

\newcommand{\p}{\partial}
\newcommand{\dd}{\mathrm{d}}

\newcommand{\uv}{{\boldsymbol u}}

\newcommand{\f}[2]{\frac{#1}{#2}}

\begin{document}

\begin{frontmatter}



\title{Spin hydrodynamics}


\author{Wojciech Florkowski}

\affiliation{organization={Institute of Theoretical Physics, Jagiellonian University},
            city={30-348 Kraków},
            country={Poland}}

\begin{abstract}
The concept of spin hydrodynamics is reexamined and briefly characterized.
\end{abstract}







\end{frontmatter}


\section{Introduction}
\label{sec:intro}

Experimental data indicating a nonvanishing spin polarization of various hadrons produced in heavy-ion collisions~\cite{STAR:2017ckg, Adam:2018ivw, Niida:2018hfw} inspired the development of spin hydrodynamics. This framework incorporates spin degrees of freedom in a standard framework of relativistic hydrodynamics~\cite{Florkowski:2017olj,Romatschke:2017ejr}.  Since spin hydrodynamics is a rather new field, different groups worldwide follow different strategies and concepts to obtain new and original results. The most popular approaches are as follows: {\bf (i)} Only the gradients of the standard hydrodynamic variables are used to determine spin polarization~\CITn{Becattini:2013fla,  Becattini:2021iol, Palermo:2024tza}. In this case, the main objects are thermal vorticity $\varpi_{\mu \nu} = -\frac{1}{2} (\p_\mu \beta_\nu-\p_\nu \beta_\mu)$ and thermal shear $\xi_{\mu \nu} = -\frac{1}{2} (\p_\mu \beta_\nu+\p_\nu \beta_\mu)$, where $\beta_\mu = u_\mu/T$, with $u_\mu$ and $T$ being the hydrodynamic flow and temperature. {\bf (ii)} The spin hydrodynamic equations are derived from the kinetic theory~\CITn{Florkowski:2017ruc, Florkowski:2017dyn, Florkowski:2018ahw, Florkowski:2018fap, Bhadury:2020puc, Bhadury:2022ulr, Weickgenannt:2019dks, Weickgenannt:2021cuo, Weickgenannt:2020aaf, Weickgenannt:2022zxs, Weickgenannt:2023nge, Wagner:2024fhf, Hu:2021pwh, Li:2020eon, Shi:2020htn}. {\bf (iii)}~The formalism is developed by making the reference to mathematically allowed forms for the energy-momentum and spin tensors, following the Israel-Stewart method~\CITn{Hattori:2019lfp, Fukushima:2020ucl, Daher:2022xon, Daher:2022wzf, Sarwar:2022yzs, Wang:2021ngp, Biswas:2023qsw, Xie:2023gbo, Daher:2024ixz, Ren:2024pur, Daher:2024bah, Gallegos:2021bzp, Hongo:2021ona, Kumar:2023ojl, She:2021lhe}. {\bf (iv)} A spin-extended Lagrangian formalism is employed~\CITn{Montenegro:2017rbu}. For a recent review see, for exmaple, \CIT{Huang:2024ffg}.

So far, very little or nothing has been done to establish relations and connections between various theoretical frameworks mentioned above. In this work we report on very recent attempts at a synthesis of the approach based on the kinetic theory with the formalism using the Israel-Stewart method, tracks {\bf (ii)} and {\bf (iii)} above.  In our opinion, by clarifying and eventually removing discrepancies between different formulations of spin hydrodynamics, we can achieve significant progress in this rapidly developing field. 

\section{Local equilibrium and generalized thermodynamic relations}
\label{sec:gtr}

One of the issues that lack consensus within the spin-hydro community is the concept of local thermodynamic equilibrium. Following our earlier works~\CITn{Florkowski:2017ruc,Florkowski:2018fap}, we suggest that local equilibrium for particles with spin corresponds to the case where the spin part of the total angular momentum is conserved. In physics terms, this corresponds to a regime where interaction is dominated by the $s$-wave scattering $(l=0)$~\CITn{Coleman:2018mew}. Conservation of the spin tensor allows, in a straightforward way, for the introduction of the tensor spin chemical potential $\Omega_{\mu\nu}$ that can be naturally interpreted as a tensor Lagrange multiplier (with $\Omega_{\mu\nu}=-\Omega_{\nu\mu}$). As a consequence, one can construct a framework of perfect spin hydrodynamics based on eleven conservation laws for the baryon number, energy, linear momentum, and spin part of the angular momentum
\bel{eq:psh}
\p_\mu N^\mu_{\rm eq}  = 0, \quad \p_\mu T^{\mu \alpha}_{\rm eq} = 0, \quad \p_\mu S^{\mu,\alpha \beta}_{\rm eq} = 0.
\eel
Here $N^\mu_{\rm eq}$ is the baryon current, $T^{\mu \alpha}_{\rm eq}$ is the energy-momentum tensor, and $S^{\mu,\alpha \beta}_{\rm eq}$ is the {\it spin tensor}.
The unknown functions to be determined by the hydrodynamic equations are: the baryon chemical potential $\mu$, temperature $T$, three independent components of the flow vector $\uv$, and the spin chemical potential $\Omega_{\mu\nu}$.

In the recent works~\CITn{Florkowski:2024bfw,Drogosz:2024gzv}, we have argued that a consistent and nontrivial structure of thermodynamic relations used in spin local equilibrium should have a tensor form:
\bel{eq:Smu_FD1}
S^\mu_{\rm eq} = -N^\mu_{\rm eq} \xi + T^{\mu \alpha}_{\rm eq}\beta_\alpha - \frac{1}{2} S^{\mu,\alpha \beta}_{\rm eq}\omega_{\alpha \beta} + \mathcal{N}^\mu,
\eel
\bel{eq:dSmu}
\dd S^{\mu}_{\rm eq}= -\xi \dd N^\mu_{\rm eq} + \beta_{\alpha}\dd T^{\mu \alpha}_{\rm eq} -  \frac{1}{2} \omega_{\alpha \beta} \dd S^{\mu,\alpha \beta}_{\rm eq},
\eel
\bel{eq:dcalN}
\dd \mathcal{N}^{\mu} = N^{\mu}_{\rm eq}\dd \xi - T^{\mu \alpha}_{\rm eq}\dd \beta_{\alpha} + \frac{1}{2} S^{\mu,\alpha \beta}_{\rm eq}\dd \omega_{\alpha \beta},
\eel
where $S^\mu_{\rm eq} $ is the entropy current. For the classical statistics, one finds that  $\mathcal{N}^{\mu} = (\cosh\xi/\sinh\xi) N^{\mu}_{\rm eq}$, where $\xi = \mu/T$. The tensor $\omega_{\alpha \beta} = \Omega_{\alpha \beta}/T$ is called the {\it spin polarization tensor}. It plays a similar role as the ratio $\xi$ -- in natural units both $\xi$ and $\omega$ are dimensionless, which allows for making expansions in $\xi$ and/or $\omega_{\alpha \beta}$. A direct consequence of \EQ{eq:dSmu} is that the conservation laws~\EQn{eq:psh} imply the entropy conservation $\p_\mu S^\mu_{\rm eq}=0$. One can also notice that~\EQ{eq:dcalN} is not independent, it follows from~\EQn{eq:Smu_FD1} and~\EQn{eq:dSmu}.

One of the main arguments for the introduction of the tensor form of the thermodynamic relations \EQSMn{eq:Smu_FD1}{eq:dcalN} is the fact that microscopic calculations indicate that the equilibrium spin tensor cannot be written in a simple phenomenological form $S^{\mu,\alpha \beta}_{\rm eq} = u^\mu S^{\alpha \beta}$, where a rank-2 antisymmetric tensor $S^{\alpha \beta}$ is interpreted as the {\it spin density tensor}~\CITn{Florkowski:2018fap}. The tensor form  \EQSMn{eq:Smu_FD1}{eq:dcalN} allows for using more complicated expressions for the spin tensor, provided other tensors are consistently derived from the same microscopic model. In practice, this means that the hydrodynamic tensors other than $S^{\mu,\alpha \beta}_{\rm eq}$ and appearing in  \EQSMn{eq:Smu_FD1}{eq:dcalN} should be expanded to the order higher by one compared to $S^{\mu,\alpha \beta}_{\rm eq}$. In this case, all the terms in~\EQSMn{eq:Smu_FD1}{eq:dcalN} are included up to the same order in the components of $\omega_{\alpha\beta}$.

 The simplest application of such a strategy is to neglect the contributions related to $S^{\mu,\alpha \beta}_{\rm eq}$ in thermodynamic relations such as~\EQSM{eq:Smu_FD1}{eq:dcalN}~\CITn{Florkowski:2018fap}. This is justified if one considers only zeroth and first order terms in the spin polarization tensor. In practice, all calculations suggest that the expansion of $S^{\mu,\alpha \beta}_{\rm eq}$ starts with terms linear in the coefficients of $\omega^{\alpha \beta}$, hence, nontrivial spin contributions in thermodynamic relations start with quadratic terms. If solely the zeroth and first order terms in $\omega_{\alpha\beta}$ are included, the thermodynamic relations  \EQSMn{eq:Smu_FD1}{eq:dcalN} are reduced to standard scalar form without spin degrees of freedom. Moreover, in this case, the spin evolution can be treated as taking place in an externally given standard hydrodynamic background. This has allowed for studies of spin evolution in simplified~\CITn{Wang:2021ngp, Xie:2023gbo,Drogosz:2024lkx} and, very recently, also realistic models of expansion~\CITn{Singh:2024cub}. 
 
 A kind of half-measure taken to describe spin is the use of the phenomenological formula $S^{\mu,\alpha \beta}_{\rm eq} = u^\mu S^{\alpha \beta}$ without any reference to the microscopic models. Since $S^{\alpha \beta}$ depends on $T$, $\mu$, $u^\mu$, and $\omega_{\alpha \beta}$, one expects that the products of $S^{\alpha \beta}$ and the spin polarization tensor are again of the second order in  $\omega_{\alpha \beta}$ and may be neglected in the thermodynamic relations. To keep them included, an additional ansatz has been adopted that  $S^{\alpha \beta}$ can be treated as the zeroth-order quantity~\CITn{Hattori:2019lfp}. This approach allows for using the spin tensor in thermodynamic relations; however, it gives rise to other problems, since the assumption $S^{\alpha \beta} = S(T,\mu)\omega^{\alpha \beta}$ leads to instabilities~\CITn{Daher:2022wzf, Sarwar:2022yzs, Xie:2023gbo, Daher:2024ixz, Ren:2024pur, Daher:2024bah}. Such instabilities do not appear if the electriclike and magneticlike components of  the spin density tensor $S^{\alpha \beta}$ depend differently on the electriclike and magneticlike components of the spin polarization tensor $\omega_{\alpha \beta}$. In fact, such a different dependence is natural in the microscopic models~\CITn{Daher:2024ixz}, hence, we arrive again at the kinetic-theory formula for the spin tensor and a set of generalized thermodynamic relations \EQSMn{eq:Smu_FD1}{eq:dcalN} as an appropriate base for construction of the perfect spin hydrodynamics.

It is important to stress at this point that the only expansion parameter discussed so far has been the magnitude of the spin polarization tensor components $\omega_{\mu\nu}$, which is a dimensionless quantity in natural units. The gradient expansion is introduced below in the context of including dissipation. 

\section{Close-to-equilibrium dynamics}
\label{sec:gtr}
 
 To go beyond the scope of the perfect spin hydrodynamics defined above, we closely follow the method of Israel and Stewart~\CITn{Israel:1979wp} and replace the equilibrium currents $N^\mu_{\rm eq}$, $T^{\mu \alpha}_{\rm eq}$ and $S^{\mu, \alpha \beta}_{\rm eq}$ in \EQ{eq:Smu_FD1} by the general nonequilibrium expressions. The latter can be written as the equilibrium terms plus nonequilibrium corrections: $N^\mu = N^\mu_{\rm eq} + \delta N^\mu$, $T^{\mu \alpha} = T^{\mu \alpha}_{\rm eq} + \delta T^{\mu \alpha}$ and $S^{\mu, \alpha \beta} = S^{\mu, \alpha \beta}_{\rm eq} +\delta S^{\mu, \alpha \beta}$. In this way we obtain
\bel{eq:HmuN}
S^\mu =  T^{\mu \alpha} \beta_\alpha-\f{1}{2} \omega_{\alpha\beta} S^{\mu, \alpha \beta}
-\xi N^\mu + {\cal N}^\mu.
\eel 
To proceed further, we calculate the entropy production $\p_\mu S^\mu$ using again the conservation laws. The main difference with respect to the perfect fluid description is that the spin tensor is not conserved now. We recall that in the general case only the total angular momentum is conserved. In this case the conservation law $\p_\mu J^{\mu, \alpha\beta}=0$ with $J^{\mu, \alpha\beta} = x^\alpha T^{\mu \beta} - x^\beta T^{\mu \alpha} + S^{\mu, \alpha\beta}$ gives $\p_\mu S^{\mu, \alpha\beta}= T^{\beta \alpha}-T^{\alpha \beta}$. The calculation of  the divergence of the entropy current defined by \EQ{eq:HmuN} gives
\bel{eq:divS}
\p_\mu S^\mu =  
- \delta N^\mu \p_\mu \xi
+ \delta T^{\mu \lambda}_s \p_\mu \beta_\lambda 
+ \delta T^{\mu \lambda}_a \LB \p_\mu \beta_\lambda 
- \omega_{\lambda\mu} \RB
-\f{1}{2} \delta S^{\mu, \alpha \beta} \p_\mu \omega_{\alpha\beta} ,
\eel
where $ \delta T^{\mu \lambda}_s$ and $ \delta T^{\mu \lambda}_a$ are symmetric and antisymmetric dissipative corrections to the energy-momentum tensor, respectively. These and other dissipative terms can be determined from the condition  $\p_\mu S^\mu \geq 0$, following the ideas originally introduced in~\CITn{Hattori:2019lfp}. This leads to expressions where $\delta N^\mu$, $\delta T^{\mu \lambda} =  \delta T^{\mu \lambda}_s +  \delta T^{\mu \lambda}_a$, and $\delta S^{\mu, \alpha \beta}$ are expressed as combinations of the gradients of the hydrodynamic variables used in the perfect-fluid description. 

The strategy outlined above clearly suggests that our approach is based on a two-fold expansion, in both the magnitude of $\omega_{\alpha\beta}$ and gradients of hydrodynamic variables. As long as we remain at the perfect-fluid level, we consider an expansion in $\omega_{\alpha\beta}$ only. If dissipation is included, we have to include in parallel the gradient expansion. This approach is similar to the treatment of the baryon chemical potential $\mu$ in standard hydrodynamics -- one can always consider an expansion in the magnitude of $\xi =\mu/T$ together with the gradient expansion. 

Although \EQ{eq:divS} or its special case for $\xi=0$ was obtained before (see, for example: Eq.~(10) in \CITn{Hattori:2019lfp}, (23) in \CITn{Fukushima:2020ucl}, (21) in \CITn{Biswas:2023qsw}, and the QFT analysis in \CITn{Becattini:2023ouz}), the previous studies considered always the local equilibrium state without the second-order corrections discussed above. Thus, it is important to extend the previous analyses by considering a different reference point for local equilibrium quantities. 

Equation \EQn{eq:divS} implies that the {\it global equilibrium} is defined by the generalized Tolman-Klein conditions~\CITn{Tolman:1934,Klein:1949}, which include the two standard equations, $\p_\mu \xi = 0$ and $\p_{( \mu} \beta_{\lambda )} = 0$, and an extra constraint that the spin polarization tensor is given by the thermal vorticity, $\omega_{\lambda\mu} = \p_{[ \mu} \beta_{\lambda ]}$.~\footnote{Here we follow the standard notation:  the round (squared) brackets denote a symmetric (antisymmetric) part of a tensor.} Nevertheless, in {\it local equilibrium} $\omega_{\lambda\mu}$ and $\p_{[ \mu} \beta_{\lambda ]}$ are not directly related and may be significantly different from each other. In this respect, we differ from the concept of local equilibrium originally proposed in~\CITn{Becattini:2013fla}. As stated above, the behavior of the spin polarization tensor, $\omega_{\mu\nu} = \Omega_{\mu\nu}/T$, is similar to the behavior of the ratio $\xi = \mu/T$ in standard (spinless) relativistic hydrodynamics. In global equilibrium $\xi =$~const., while in local equilibrium a direct connection between $T$ and $\mu$ is lost. 

\section{Summary}

Following Refs.~\CITn{Florkowski:2024bfw, Drogosz:2024gzv}, we propose to use a hybrid form of spin hydrodynamics that combines, in our opinion, the most attractive features of various approaches to spin hydrodynamics that have been developed so far. The hybrid approach allows  to establish relations and connections between such approaches and offers solutions to several important problems:

\medskip \noindent
{\bf  (i)} Expansion up to the second order in the spin polarization tensor $\omega_{\alpha\beta}$ allows us to maintain nontrivial spin thermodynamic relations without the contradictory assumption that the spin tensor and the spin polarization tensor are of different orders. 

\medskip \noindent
{\bf( ii)} The problems with stability and causality found in the frameworks using the Israel-Stewart method can be solved by the reference to the kinetic-theory result that explains different dependence of the spin tensor on the electriclike and magneticlike components~\CITn{Daher:2024ixz}. 

\medskip \noindent
{\bf  (iii)} The two-fold character of the expansion described above allows to disentangle different phenomena that otherwise seem to be connected. For example, since in the dissipative spin hydrodynamics one finds the combination $\beta a^\mu + \beta^2 \nabla^\mu T - 2 \omega^{\mu\nu} u_\nu$ (where $\beta$ is the inverse temperature, $a^\mu$ the four-acceleration, and $\nabla^\mu$ the transverse gradient), it is popular in the literature to assume that $\omega_{\mu\nu}$ has the character of a gradient term. There is no reason to do so in our framework. Note that $\beta a^\mu + \beta^2 \nabla^\mu T =0$ for the Bjorken expansion, where there is no reason to assume that  $\omega_{\mu\nu}$ vanishes. 

\medskip
Our approach defines local equilibrium for particles with spin in simple physical terms as the phase dominated by the $s$-wave scattering.The inclusion of dissipation relaxes this assumption and allows for processes that change the orbital and spin parts of the total angular momentum. Certainly, it remains to be seen whether such a scheme turns out to be useful and reliable in physical applications to interpret the data. The first encouraging steps in this direction have been taken in~\CIT{Singh:2024cub}.

\bigskip
\noindent
{\it Acknowledgements.}  I thank my collaborators Zbigniew Drogosz and Mykhailo Hontarenko for fruitful common studies of the problems presented in this contribution. I also thank Valeriya Mykhaylova for critical comments regarding the manuscript.This work was supported in part by the Polish National Science Centre (NCN) Grant No. 2022/47/B/ST2/01372.

\bibliographystyle{elsarticle-num-names} 
\bibliography{boost-inv-lit}



\end{document}